\documentclass[a4paper]{jpconf}
\usepackage{graphicx}

\newcommand{\eq}[1]{\begin{equation} #1 \end{equation}}

\begin{document}
\title{Analysis of hadron yield data within hadron resonance gas model with multi-component eigenvolume corrections}

\author{Volodymyr Vovchenko$^{1,2,3}$, Horst Stoecker$^{1,2,4}$}

\address{$^1$Frankfurt Institute for Advanced Studies, Goethe Universit\"at Frankfurt, D-60438 Frankfurt am Main, Germany}
\address{$^2$Institut f\"ur Theoretische Physik,
Goethe Universit\"at Frankfurt, D-60438 Frankfurt am Main, Germany}
\address{$^3$Department of Physics, Taras Shevchenko National University of Kiev, 03022 Kiev, Ukraine}
\address{$^4$GSI Helmholtzzentrum f\"ur Schwerionenforschung GmbH, D-64291 Darmstadt, Germany}

\ead{vovchenko@fias.uni-frankfurt.de}

\begin{abstract}
We analyze the sensitivity of thermal fits to heavy-ion hadron yield data of ALICE and NA49 collaborations to the systematic uncertainties in the hadron resonance gas (HRG) model related to the modeling of the eigenvolume interactions.
We find a surprisingly large sensitivity in extraction of chemical freeze-out parameters to the assumptions regarding eigenvolumes of different hadrons.
We additionally study the effect of including yields of light nuclei into the thermal fits to LHC data and find even larger sensitivity to the modeling of their eigenvolumes. The inclusion of light nuclei yields, thus, may lead to further destabilization of thermal fits. Our results show that modeling of eigenvolume interactions plays a crucial role in thermodynamics of HRG and that conclusions based on a non-interacting HRG are not unique.
\end{abstract}

\section{Introduction}
Thermodynamic models have long been employed to estimate the temperatures reached in the relativistic heavy-ion collisions~\cite{Stoecker:1981za,Molitoris:1985wa,Stoecker:1986ci,Hahn:1987tz}.
Particularly the  hadron-resonance gas (HRG) model has been quite successful in describing the hadron multiplicities data in relativistic nucleus-nucleus 
collisions~\cite{Cleymans:1992zc,Andronic:2005yp}, possibly with additional introduction of parameters to regulate deviations from full chemical equilibrium~\cite{Letessier:2005qe,Becattini:2005xt,Petran:2013lja}.

In its simplest version the HRG is described as a gas of non-interacting hadrons and resonances.
In a more realistic HRG model one has to take into account the attractive and repulsive interactions between hadrons.
In this work we study the influence of short-range repulsive interactions between hadrons on thermal fits to hadron yield data.

\section{Model description}
The repulsive interactions between hadrons are modeled via eigenvolume (EV) correction of the van der Waals type
\cite{Hagedorn:1980kb,Gorenstein:1981fa,Kapusta:1982qd,Rischke:1991ke}.
For a multi-component case there is more than a single way to model the eigevolume corrections.
In our analysis we use two different formulations:
the ``Diagonal'' eigenvolume HRG model from~\cite{Yen:1997rv}, and the ``Crossterms'' EV model introduced in \cite{Gorenstein:1999ce}
~(see Ref.~\cite{Vovchenko:2016ebv} for technical details).
The ``Diagonal'' model is simpler but it is not consistent with the 2nd order virial expansion of the equation of state of hard spheres. The ``Crossterms'' model, on the other hand, while technically more involved, is consistent with that expansion.

In our analysis we include the established strange and non-strange hadrons listed in the Particle Data Tables~\cite{Agashe:2014kda}, along with their decay branching ratios. This includes mesons up to $f_2(2340)$ and (anti)baryons up to $N(2600)$.
The finite width of the resonances is taken into account by adding the additional integration over their Breit-Wigner shapes.
The feed-down from strong and electromagnetic decays of the unstable resonances to the total hadron yields is included in the standard way. 
The full chemical equilibrium assumed in this work, i.e. we do not consider here possible under- or over-saturation of the light and/or strange quarks which improves the data description but also introduces additional parameters~(see, e.g., Ref.~\cite{Letessier:2005qe}).

\section{Mass-proportional eigenvolume at finite $\mu_B$}
First
a bag-model inspired parametrization with the hadron eigenvolume proportional to its mass through a bag-like constant is employed, i.e. 
\eq{\label{eq:BagEV}
v_i = m_i / \varepsilon_0.
}
Such eigenvolume parametrization had been obtained for the heavy Hagedorn resonances, and was used to describe their thermodynamics~\cite{Hagedorn:1980kb,Kapusta:1982qd}, their
effect on particle yield ratios~\cite{NoronhaHostler:2010yc},
and compared to lattice QCD~\cite{Albright:2014gva}.
The EV-HRG model fits are using the hadron yield
data of the NA49 collaboration,
which include
$4\pi$ yields of charged pions, charged kaons, $\Xi^-$, $\Xi^+$, $\Lambda$, $\phi$, and, if available, $\Omega$, $\bar{\Omega}$, measured in most central Pb+Pb collisions at $\sqrt{s_{\rm NN}} = 6.3, 7.6, 8.8, 12.3$, 
and $17.3$~GeV~\cite{NA49data-1a,NA49data-1b,NA49data-1c,NA49data-2a,NA49data-2b,NA49data-2c,NA49data-2d,NA49data-2e}. 
Additionally, the data on the total number of participants, $N_W$, is identified with total net baryon number and is included in the fit.

\begin{figure}[!h]
\centering
\includegraphics[width=0.49\textwidth]{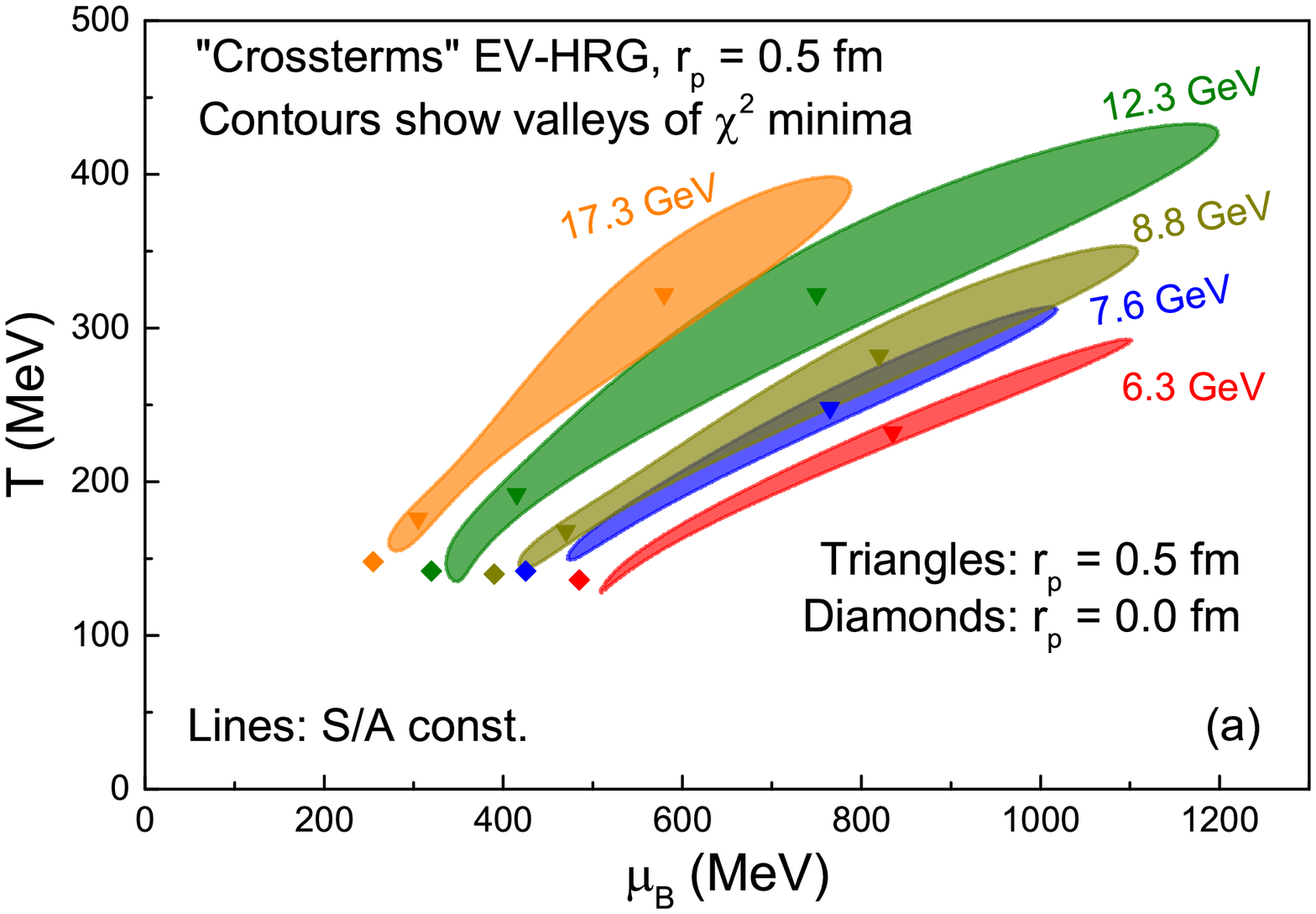}
\includegraphics[width=0.49\textwidth]{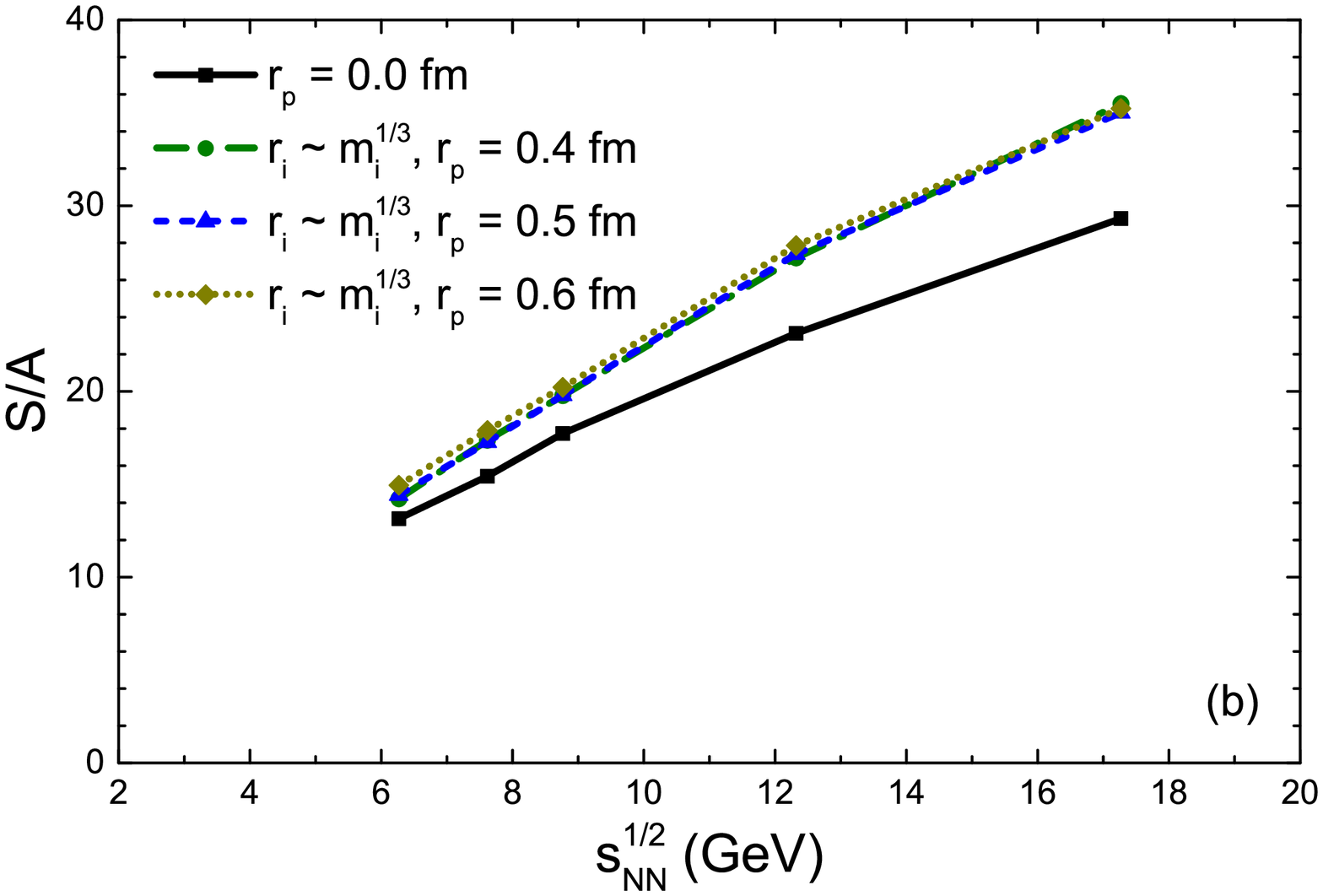}
\caption[]{
(a) Regions in the $T$-$\mu_B$ plane where the ``Crossterms'' eigenvolume HRG model with bag-like constant $\varepsilon_0$ fixed to reproduce the hard-core radius of $r_p = 0.5$~fm yields a better fit to the NA49 data as compared to the point-particle HRG model.
The solid lines show the isentropic curves for the eigenvolume model, which go through the global $\chi^2$ minima.
(b) Collision energy dependence of entropy per baryon at global minima of thermal fits to NA49 data.
}\label{fig:chi2-Tmu}
\end{figure}

For illustration purposes we consider here only the ``Crossterms'' model and we fix the $\varepsilon_0$ to reproduce the hard-core proton radius of $r_p = 0.5$~fm. 
A much more detailed study regarding fits with mass-proportional eigenvolumes can be found in~\cite{Vovchenko:2016ebv}. Figure $\ref{fig:chi2-Tmu}$a depicts the regions in the $T$-$\mu_B$ plane where the fit quality
of NA49 data in the EV model is better than in the non-interacting HRG.
The location of the fit minima within non-interacting HRG are shown by diamonds, and are consistent with the systematics established in numerous previous studies.
At all five NA49 energies wide regions of improved $\chi^2$ values are observed at high temperatures and chemical potentials. 
At given bombarding energy, for a given set of radii, the EV HRG model fits do not just yield a single $T-\mu_B$ pair, but a whole range of $T-\mu_B$ pairs, each with similarly good fit quality. These pairs form a valley in the $T-\mu_B$ plane along a line of nearly constant entropy per baryon, $S/A$.
We also show the energy dependence of values of  $S/A$ (Fig. $\ref{fig:chi2-Tmu}$b). The $S/A$, extracted at different energies, is a robust observable: it is almost independent of the details of the modeling of the EV interactions and of the specific $T-\mu_B$ values obtained.

\section{Two-component model and role of light nuclei}
\begin{figure}[!h]
\centering
\includegraphics[width=0.49\textwidth]{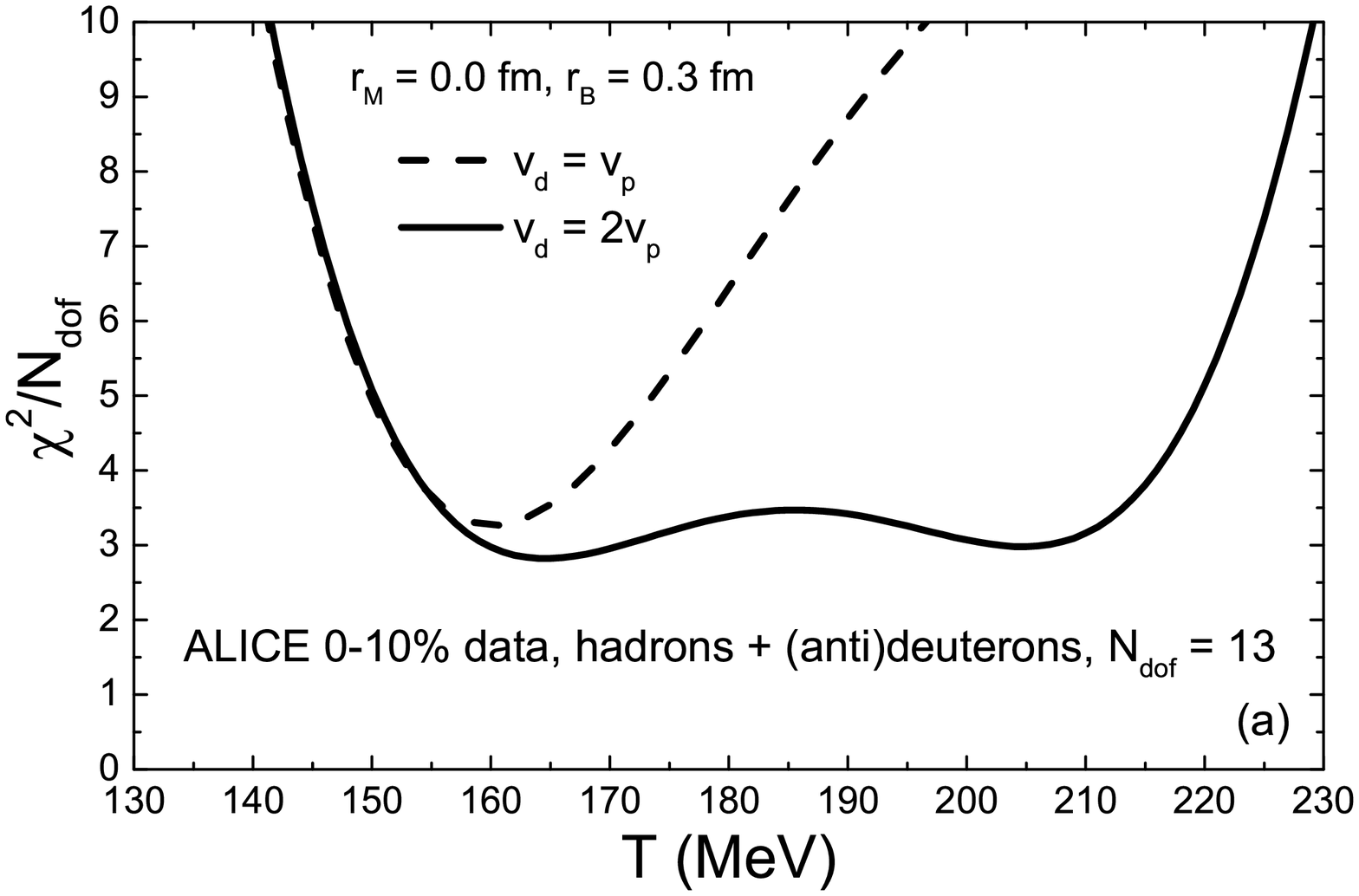}
\includegraphics[width=0.49\textwidth]{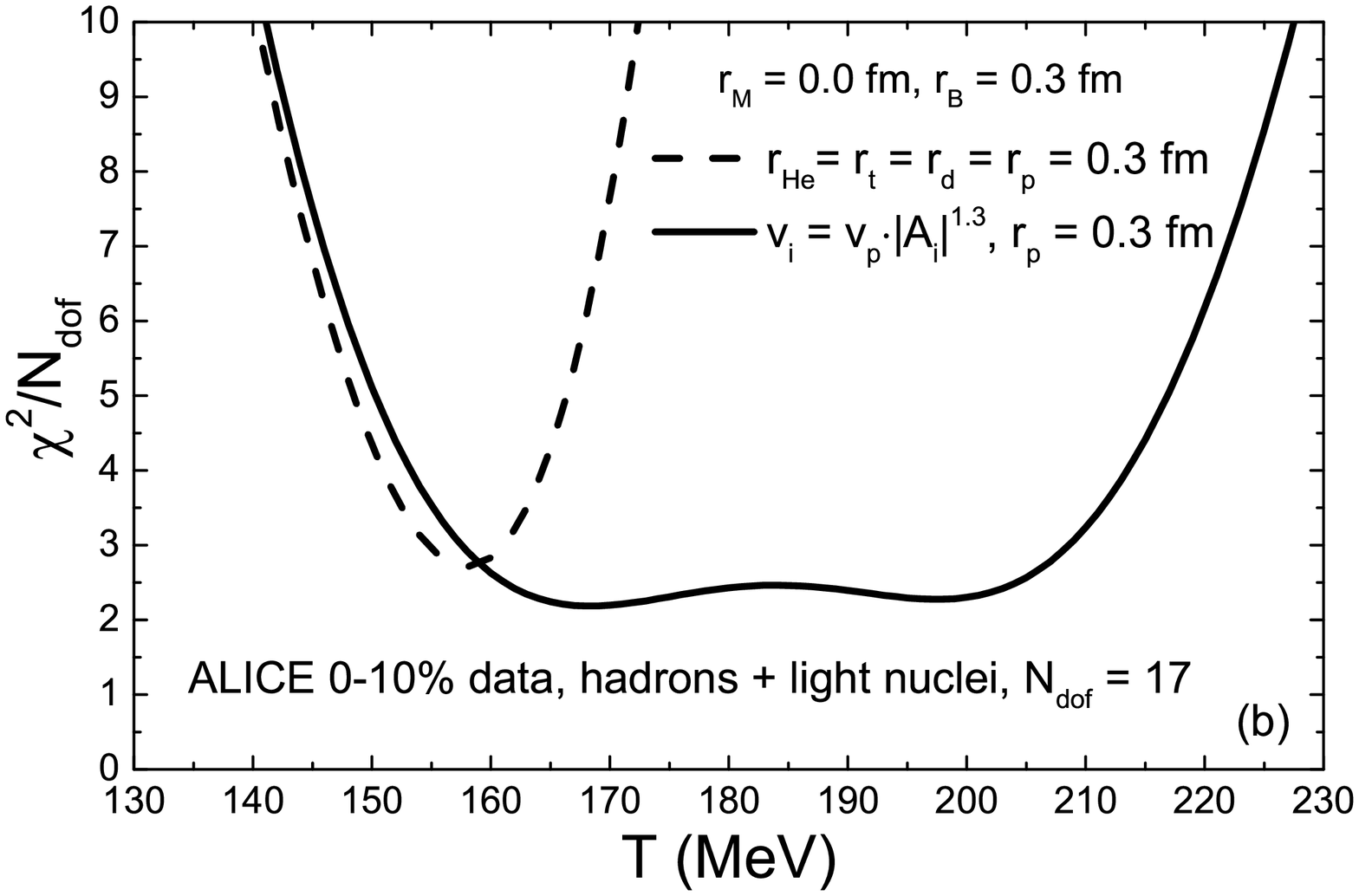}
\caption[]{
The temperature dependence of $\chi^2 / N_{\rm dof}$ of fit to ALICE data on hadron + light nuclei yields in 0-10\% most central Pb+Pb collisions at 2.76~TeV within eigenvolume HRG model with point-like mesons, baryons with hard-core radius of 0.3~fm, and different assumptions for eigenvolumes of light nuclei.
In (a) only (anti)deuterons are added to fit with $r_d = r_p = 0.30$~fm (dashed line)  and $r_d = 2^{1/3} \, r_p \simeq 0.38$~fm (solid line).
In (b) light nuclei up to $^3_{\Lambda}$H are added with $r_{\rm He} = r_t = r_d = r_p = 0.3$~fm~(dashed), and $v_i = v_p \cdot |A_i|^{1.3}$~(solid),
where $|A_i|$ is mass number.
}\label{fig:deuteron}
\end{figure}

We also consider parameterization where all mesons are assumed to be point-like and where all baryons have a fixed hard-core radius $r_p = 0.3$~fm. 
As demonstrated in \cite{Andronic:2012ut} this parameterization yields a good description of lattice data on the pressure at least up to $T=190$~MeV. 
It was also shown in~\cite{Vovchenko:2015cbk} that it leads to an irregular $\chi^2$ profile of thermal fit to ALICE hadron yield data  with wide double minimum structure in the 155-210~MeV temperature range.
Here we investigate how inclusion of light nuclei influences the fit.
For that we consider the 0-10\% centrality ALICE data which 
has a rich amount of data regarding yields of light nuclei.
The actual data used for fitting includes midrapidity yields of charged pions, charged kaons, and (anti)protons~\cite{Abelev:2013vea}, 
(anti)$\Xi^-$ and 
(anti)$\Omega$~\cite{ABELEV:2013zaa},
(anti)$\Lambda$ and $K^0_S$~\cite{Abelev:2013xaa,Schuchmann:2015lay},
$\phi$~\cite{Abelev:2014uua},
(anti)deuterons and (anti)$^3$He~\cite{Adam:2015vda}, 
and (anti)$^3_{\Lambda}$H~\cite{Adam:2015yta}.
The addition of light nuclei requires additional assumptions regarding their eigenvolumes. 
First we consider only addition of (anti)deuterons and compare two cases: (1) deuteron has same eigenvolume as baryons, i.e. $r_p = r_d = 0.3$~fm; (2) deuteron has a twice larger eigenvolume compared to baryons, i.e. $r_d \simeq 0.38$~fm. The results of the calculations of $\chi^2$ temperature profile are depicted in Fig.~\ref{fig:deuteron}a. Note that $\mu_B$ is also fitted at each $T$ and found to be consistent with zero within fit errors.

If deuterons are assumed to have same eigenvolume as protons then the $\chi^2$ profile has a regular structure with a minimum at $T \simeq 161$~MeV. Thus, even though inclusion of light nuclei may be questionable in general, it would seem that inclusion of deuterons into thermal fit would stabilise them with regards to the modeling of eigenvolume corrections. This conclusion, however, is illusory. 
Changing deuteron eigenvolume to a physically more motivated value equal to twice that of proton one gets a very different $\chi^2$ profile: a two-minimum structure in a wide 155-210~MeV temperature range, with improved fit quality at global minimum. This change is attributed to a larger suppression of (anti)deuteron yield at higher temperatures due to larger eigenvolume.
It is remarkable that fit can be so sensitive to the properties of only single particle species. Similarly, the inclusion of yields of $^3$He and $^3_{\Lambda}$H does not help to stabilise the fit, as seen in Fig.~\ref{fig:deuteron}b. This is so despite the remarkably large number of degrees of freedom $N_{\rm dof} = 17$.
We therefore conclude that introduction of light nuclei into thermal fits leads to a further destabilisation of the fit as it requires non-trivial assumptions regarding their eigenvolumes.

\section{Conclusions}
It is shown that thermal fits are very sensitive to the details of the modeling of the eigenvolume interactions. At given collision energy, the EV HRG model fits do not just yield a single $T-\mu_B$ pair, but a whole range of $T-\mu_B$ pairs, each with similarly good fit quality. The entropy per baryon, on the other hand, appears to be a robust observable: its extracted value depends weakly on the modeling of hadron eigenvolumes. Finally, we explore the effect of including light nuclei into thermal fits and, somewhat surprisingly, find that results are even more sensitive to the assumptions regarding their eigenvolume parameters.

\end{document}